\documentstyle[epsfig]{elsart}

\begin{document}
\begin{frontmatter}
\title{Right-handed Electrons in Radiative Muon Decay}

\author{L.~M.~Sehgal}
\address{Institute of Theoretical Physics (E), RWTH Aachen, D-52056 Aachen, Germany}

\begin{abstract}
Electrons emitted in the radiative decay 
$\mu^- \to e^- \bar{\nu}_e \nu_\mu \gamma$ 
have a significant probability of being right-handed, even in the limit 
$m_e \to 0$. 
Such ``wrong-helicity'' electrons, 
arising from helicity-flip bremsstrahlung, contribute an amount 
$\frac{\alpha}{4 \pi} \Gamma_0$
to the muon decay width 
($\Gamma_0 \equiv G_F^2 m_\mu^5 / (192 \pi^3)$).
We use the helicity-flip splitting function $D_{hf}(z)$ 
of Falk and Sehgal (Phys.\ Lett.\ B {\bf 325}, 509 (1994)) to obtain the spectrum 
of the right-handed electrons and the photons that accompany them. 
For a minimum photon energy $E_\gamma = 10 \, MeV$ $(20 \, MeV)$, 
approximately $4 \%$ ($7 \%$) of electrons in radiative $\mu$-decay are right-handed.
\end{abstract}
\end{frontmatter}

It is usually thought that in V-A theory, electrons emitted in muon decay 
are purely left-handed, in the limit $m_e \to 0$. 
This statement, however, is not true for electrons in the radiative decay 
$\mu^- \to e^- \bar{\nu}_e \nu_\mu \gamma$, 
where the photon is the result of inner bremsstrahlung. 
We show in this Letter that radiative muon decay contains a well-defined constituency 
of right-handed electrons, contributing an amount 
$\frac{\alpha}{4\pi} \Gamma_0$ ($\Gamma_0 \equiv G_F^2 m_\mu^5 / (192 \pi^3)$) 
to the decay width. We calculate the spectrum of these ``wrong-helicity'' electrons, 
and of the photons that accompany them. These spectra are compared with the 
unpolarized spectra, summed over electron helicities. 
This comparison provides a quantitative measure of the right-handed 
fraction and its distribution in phase space.

The appearance of ``wrong-helicity'' electrons in the decay 
$\mu^- \to e^- \bar{\nu}_e \nu_\mu \gamma$, 
even in the limit $m_e \to 0$, is a consequence of helicity-flip bremsstrahlung in 
quantum electrodynamics, a feature first noted by Lee and Nauenberg \cite{Lee:is}. 
It was found that in the radiative scattering of electrons by a Coulomb field, 
the probability of helicity-flip 
(i.e. $e_L \to e_R$ or $e_R \to e_L$) 
did not vanish in the limit 
$m_e \to 0$. This unexpected result, in apparent contradiction to the naive expectation of 
helicity conservation in the $m_e \to 0$ limit, arises from the fact, 
that the helicity-flip cross section for bremsstrahlung at small 
angles has the form 
\begin{equation}
\frac{d\sigma}{d\theta^2} \sim \frac{ \left( \frac{m_e}{E_e} \right) ^2}
{\left( \left( \frac{m_e}{E_e} \right) ^2+\theta^2 \right)^2},
\end{equation}
which, when integrated over angles, gives a finite non-zero answer 
in the limit $m_e \to 0$.

In Ref. \cite{Falk:1993tf}, Falk and Sehgal examined the helicity structure 
of the brems\-strah\-lung process in an equivalent particle approach, 
and showed that helicity-flip radiation 
$e_L^- \to e_R^- + \gamma(z)$, in the limit $m_e \to 0$, 
can be described by a simple and universal splitting (or fragmentation) function
\begin{equation}
D_{hf}(z)=\frac{\alpha}{2 \pi} \, z,
\end{equation}
where $z={E_\gamma}/{E_e}$ is the ratio of the photon energy to the energy 
of the radiating electron. This function is analogous to the familiar Weizs\"acker-Williams 
function describing helicity-conserving (non-flip) bremsstrahlung 
\begin{equation}
D_{nf}(z)=\frac{\alpha}{\pi} \, \, \frac{1+(1-z)^2}{z} \, \log \left( \frac{E_e}{m_e} \right). 
\end{equation}
Several applications of the helicity-flip function $D_{hf}(z)$ were considered in 
\cite{Falk:1993tf}, including the process 
$e_R^- + p \to \nu_L + \gamma + X$
(``fake right-handed currents'') and 
$e_\lambda^- e_\lambda^+ \to f \bar{f} \gamma$ 
(wrong-helicity $e^+ e^-$ annihilation). 
It was shown that the splitting function approach reproduced the results of the usual bremsstrahlung 
calculation in which the limit $m_e \to 0$ was taken at the end \cite{Contopanagos:ga,Jadach:1987ws}. 
Subsequently, the equivalent-particle technique has been successfully applied to other 
he\-li\-ci\-ty-flip processes such as 
$\pi^- \to e_L^- \bar{\nu} \gamma$ and $Z^0 \to e_L^- e_L^+ \gamma$ 
\cite{Trentadue_Smilga}.

Recently, in an analysis of radiative corrections to the electron spectrum in muon decay, 
$\mu^- \to e^- \bar{\nu}_e \nu_\mu$, 
Fi\-scher et al. \cite{Fischer:2002hn} have noted that the radiative correction to the 
helicity of the electron, calculated in an early paper by Fi\-scher and Scheck 
\cite{Fischer:zh},can be reproduced in a simple way using the helicity-flip function 
$D_{hf}(z)$. This has motivated us to examine the helicity-dependence of the radiative decay 
$\mu^- \to e^- \bar{\nu}_e \nu_\mu \gamma$, 
to determine the incidence and spectrum of wrong-helicity (right-handed) electrons in this channel.

The electron spectrum in ordinary (non-radiative) muon decay 
$\mu^- \to e^- \bar{\nu}_e \nu_\mu$ 
has, in Born appoximation, the well-known form
\begin{equation}
\left( \frac{d\Gamma}{dx \, d\cos\theta_e} \right)^{\rm non-rad} =
\Gamma_0 \, [x^2(3-2x)+x^2(1-2x)\cos (\theta_e)] 
\label{e-spectrum-non-rad},
\end{equation}
where 
$x = 2 E_e/m_\mu$ and $\theta_e$ 
is the angle of the electron relative to the spin of the muon. 
We can obtain from this the spectrum of the radiative channel 
$\mu^- \to e^- \bar{\nu}_e \nu_\mu \gamma$, 
using the splitting functions $D_{hf}$ or $D_{nf}$. 
In the specific case of right-handed electrons in the final state, the spectrum, 
in the limit $m_e \to 0$ (collinear bremsstrahlung), is given by

\begin{eqnarray} 
& & \left( \frac{d\Gamma}
{dx_e \, d\cos\theta_e} \right)_{e_R^-}^{\rm rad} = \nonumber \\ 
& & \int\limits_{0}^{1}  dx \int\limits_{0}^{1}dz 
\left( \frac{d\Gamma}{dx \, d\cos\theta_e} \right)^{\rm non-rad}
 D_{hf}(z) \, \delta(x_e-x(1-z)) \, \theta(xz-x_{\gamma 0}), 
\end{eqnarray}

where the $\theta$-function in the integrand has been inserted to allow for 
a minimum energy cut on the photon:
\begin{equation}
x_\gamma \equiv \frac{2E_\gamma}{m_\mu} \ge x_{\gamma 0}. 
\end{equation}
The result of the integration is
\begin{equation} 
\left( \frac{d\Gamma}{dx_e \, d\cos\theta_e} \right)_{e_R^-}^{\rm rad} =
\Gamma_0 \frac{\alpha}{2 \pi}[A(x_e,x_{\gamma 0})+\cos (\theta_e) B(x_e,x_{\gamma 0})]
\label{e-spectrum-cutoff},
\end{equation}
where
\begin{eqnarray} 
A(x_e,x_{\gamma 0}) = & - & \frac{2}{3} [1-(x_e + x_{\gamma 0})^3] 
                          + \frac{1}{2}[1-(x_e + x_{\gamma 0})^2](2x_e+3) \nonumber \\
                    & - & 3x_e[1-(x_e + x_{\gamma 0})], \nonumber \\
B(x_e,x_{\gamma 0}) = & - & \frac{2}{3} [1-(x_e + x_{\gamma 0})^3] 
                          + \frac{1}{2}(1+2x_e)[1-(x_e+x_{\gamma 0})^2] \\
                    & - & x_e[1-(x_e+x_{\gamma 0})]. \nonumber
\end{eqnarray}
Integrating over 
$\cos \theta_e$ and $x_e$ ($0 \le x_e \le 1-x_{\gamma 0}$), 
we obtain 
\begin{equation}
\Gamma_{e_R^-}^{\rm rad} (x_{\gamma 0}) =
\Gamma_0 \, \frac{\alpha}{\pi} \, [\frac{1}{4} - x_{\gamma 0}^2 
+ x_{\gamma 0}^3 - \frac{1}{4} x_{\gamma 0}^4].
\label{Gamma-cutoff} 
\end{equation}
\bigskip
If no cut is imposed on the photon energy (i.e. $x_{\gamma 0}=0$) 
the spectrum of right-handed electrons given in Eq.(\ref{e-spectrum-cutoff}) reduces to
\begin{equation} 
\left( \frac{d\Gamma}{dx_e \, d\cos\theta_e} \right)_{e_R^-}^{\rm rad} =
\Gamma_0 \, \frac{\alpha}{2 \pi} \, \frac{1}{6} (1-x_e)^2[(5-2x_e)-\cos(\theta_e) (2x_e+1)],
\label{e-spectrum-no-cutoff}
\end{equation}
which coincides with the result obtained by Fi\-scher and Scheck \cite{Fischer:zh}.

The helicity-flip fragmentation function also gives a simple way of 
calculating the spectrum of photons accompanying right-handed electrons in
$\mu^- \to e^- \bar{\nu}_e \nu_\mu \gamma$.
In the collinear limit ($m_e \to 0$), we have 
\begin{eqnarray} 
\left( \frac{d\Gamma}{dx_\gamma \, d\cos\theta_\gamma} \right)_{e_R^-}^{\rm rad} & = &
\int\limits_{0}^{1}  dx \int\limits_{0}^{1}dz 
\left( \frac{d\Gamma}{dx \, d\cos\theta_e} \right)_{\theta_e=\theta_\gamma}^{\rm non-rad}
 D_{hf}(z) \, \delta(x_\gamma-xz)\nonumber \\
& = & \Gamma_0 \, \frac{\alpha}{2\pi} \, x_\gamma(1-x_\gamma)[(2-x_\gamma)-x_\gamma \cos(\theta_\gamma)].
\label{photon-spectrum}
\end{eqnarray}
Integrating over all photon energies and over 
$\cos \theta_\gamma$ 
we get 
$\Gamma_{e_R^-}^{\rm rad} = \frac{\alpha}{4\pi}\Gamma_0$,
which is the same as Eq.(\ref{Gamma-cutoff}) for $x_{\gamma 0}=0$.

\begin{figure}
\begin{center}
\unitlength1truecm
\begin{picture}(8,12)
\put(0,.5){\psfig{file=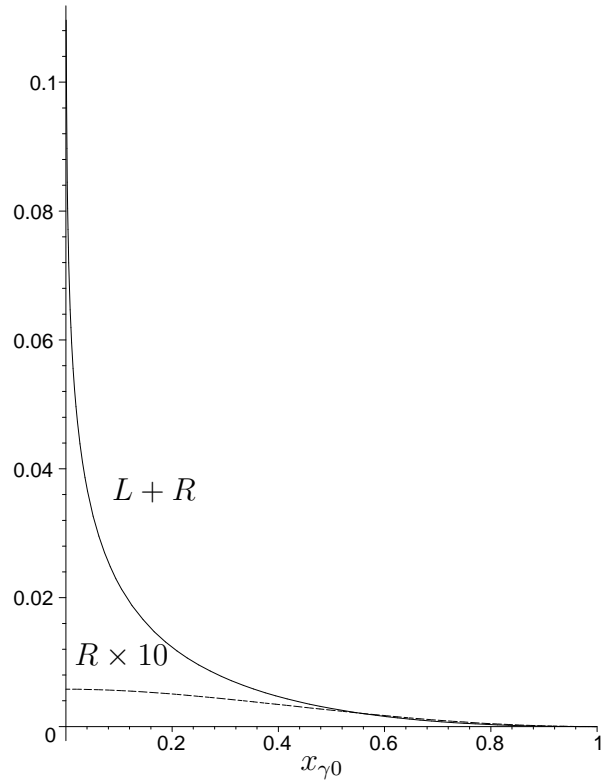,width=8cm}}
\put(4,0.9){$x_{\gamma 0}$}
\put(1.5,4.5){$\small{L+R}$}
\put(1,2.3){$\small{R \times 10}$}
\end{picture}
\end{center}
\caption{Radiative decay width 
$\Gamma_{e_L^- + e_R^-}^{\rm rad} (x_{\gamma 0})$ (full line),
compared with the right\-\-handed decay width 
$\Gamma_{e_R^-}^{\rm rad}(x_{\gamma 0})$ (multiplied by factor 10, dashed line)
as function of minimum photon energy $x_{\gamma 0}$. Decay
widths in units of $\Gamma_0$. \label{plot1}}
\end{figure}

\begin{figure}
\begin{center}
\unitlength1truecm
\begin{picture}(8,12)
\put(0,.5){\psfig{file=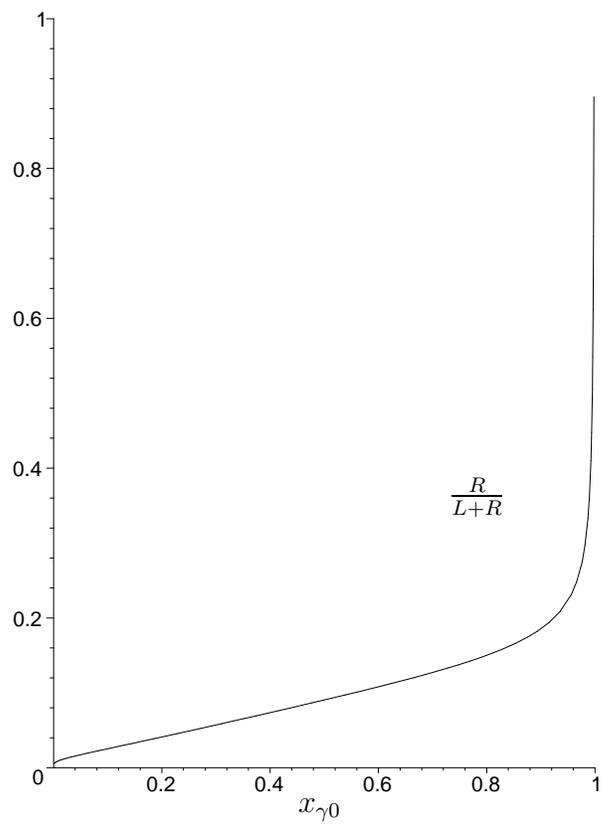,width=8cm}}
\put(4,0.9){$x_{\gamma 0}$}
\put(6,5){$\Large{\frac{R}{L+R}}$}
\end{picture}
\end{center}
\caption{Right-handed fraction $\Gamma_{e_R^-}^{\rm rad} / \Gamma_{e_R^- + e_L^-}^{\rm rad}$ 
as function of minimum photon energy $x_{\gamma 0}$. \label{plot2}} 

\end{figure}

The decay width into right-handed electrons, for a given minimum energy 
$x_{\gamma 0}$ (Eq.(\ref{Gamma-cutoff})), 
can be compared with the width summed over electron helicities. The helicity-summed photon spectrum in 
$\mu^- \to e^- \bar{\nu}_e \nu_\mu \gamma$ 
was calculated by Kinoshita and Sirlin \cite{Kinoshita:1958ru} and Eckstein and Pratt \cite{Eckstein}, 
and the integrated width, for $x_\gamma > x_{\gamma 0}$ is \cite{Eckstein}
\begin{eqnarray} 
\Gamma_{e_L^- + e_R^-}^{\rm rad} (x_{\gamma 0}) & = &
\Gamma_0 \, \frac{2\alpha}{\pi}
\left\{ \left(-\frac{17}{12}+\log(\frac{m_\mu}{m_e}) \right)
\log (\frac{1}{x_{\gamma 0}}) \right. \nonumber \\
& & - \, \frac{1}{2}(1-x_{\gamma 0})\left[ \frac{1}{6}(1-x_{\gamma 0})^3+1 \right] 
\log \left[ \frac{m_\mu^2}{m_e^2}(1-x_{\gamma 0}) \right] \nonumber \\
& & + \, \frac{1-x_{\gamma 0}}{288} 
(601-159 \, x_{\gamma 0}+ 171 \, {x_{\gamma 0}}^2 - 61 \, {x_{\gamma 0}}^3) \nonumber \\
& & \left. -\frac{\pi^2}{12}+\frac{1}{2} \sum_{n=1}^{\infty} \frac{(x_{\gamma 0})^n}{n^2} \right\}.
\label{total-width}                
\end{eqnarray}
This function is plotted in Fig.(\ref{plot1}) and compared with the right-handed width 
$\Gamma_{e_R^-}^{\rm rad}(x_{\gamma 0})$
calculated in Eq.(\ref{Gamma-cutoff}). The right-handed fraction 
$\Gamma_{e_R^-}^{\rm rad} / \Gamma_{e_R^- + e_L^-}^{\rm rad}$
is shown in Fig.(\ref{plot2}), as a function of $x_{\gamma 0}$. 
For a photon energy cut 
$E_\gamma > 10 MeV$ $(20 \, MeV)$,
this fraction is approximately 
$4 \%$ $(7 \%)$.
[It may be noted here that the branching ratio of
$\mu^- \to e^- \bar{\nu}_e \nu_\mu \gamma$,
summed over electron spins, with a photon energy cut
$E_\gamma > 10 \, MeV$, was measured in Ref. \cite{Crittenden} to be 
$(1.4 \pm 0.4) \%$.
The theoretical expression Eq.(\ref{total-width}) yields for this quantity the value 1.3\%].

A complete analysis of the channel
$\mu^- \to e^- \bar{\nu}_e \nu_\mu \gamma$
involves a study of the decay intensity in all kinematical variables. 
A variable of particular interest is the angle $\theta_{e \gamma}$
between the electron and the photon. For helicity-flip radiation, 
the characteristic angular distribution is \cite{Falk:1993tf}
\begin{equation}
\frac{dD_{hf}(z, \theta^2)}{d\theta^2} \approx \frac{\alpha}{2\pi} z 
\left( \frac{m_e}{E_e} \right)^2 
\frac{1}{\left[ \theta^2+ \left( \frac{m_e}{E_e} \right)^2 \right]^2} \, , \nonumber
\end{equation}
which is maximum at $\theta=0$ (forward direction).
By contrast, the he\-li\-ci\-ty-conser\-ving bremsstrahlung has the 
spectrum \cite{Falk:1993tf}
\begin{equation}
\frac{dD_{nf}(z, \theta^2)}{d\theta^2} \approx \frac{\alpha}{2\pi}
\frac{1+(1-z)^2}{z} 
\frac{\theta^2}{\left[ \theta^2+ \left( \frac{m_e}{E_e} \right)^2 \right]^2} \, , \nonumber 
\end{equation}
which peaks at 
$\theta \approx m_e^2/E_e^2$.
This suggests that in the decay
$\mu^- \to e^- \bar{\nu}_e \nu_\mu \gamma$,
the distribution in the angle $\theta_{e \gamma}$
between the electron and photon could be a useful discriminant 
in separating the two electron helicities. A full analysis of
the helicity-dependent decay spectrum in different kinematical 
variables will be reported elsewhere.

Summary:

\renewcommand{\theenumi}{\roman{enumi}}
\begin{enumerate}
\item The decay 
$\mu^- \to e^- \bar{\nu}_e \nu_\mu \gamma$
contains in the final state a constituency of right-handed electrons, 
which contribute an amount $\frac{\alpha}{4\pi}\Gamma_0$
to the decay width, in the limit $m_e \to 0$.

\item The spectrum of the right-handed electrons 
is given by Eq.(\ref{e-spectrum-cutoff}),
and reduces to Eq.(\ref{e-spectrum-no-cutoff}) if no cut on
photon energy is imposed. The latter differs in a characteristic way
from the spectrum of left-handed electrons, which (on account
of the soft $1/x_\gamma$ nature of helicity-conserving bremsstrahlung)
tends to follow the non-radiative pattern Eq.(\ref{e-spectrum-non-rad}).
Thus the energy spectra, integrated over angles are
$(d\Gamma/dx_e)_R \sim (1-x_e)^2(5-2x_e), \,
(d\Gamma/dx_e)_L \sim x_e^2(3-2x_e)$,
while the angular distribution, integrated over energies, is
$(d\Gamma/d\cos \theta_e)_{L, R} \sim (1-\frac{1}{3} \cos \theta_e)$,
the same for $e_L^-$ and $e_R^-$.

\item The photon spectrum associated with right-han\-ded elec\-trons is \\
$(d\Gamma/dx_\gamma)_R \sim x_\gamma (1-x_\gamma)(2-x_\gamma)$,
and is hard compared to that accompanying left-handed electrons
$(d\Gamma/dx_\gamma)_L \sim 1/x_\gamma$.

\item The right-handed fraction 
$\Gamma_{e_R^-} / (\Gamma_{e_R^-} + \Gamma_{e_L^-})$
has been calculated as a function of the photon energy cut 
$x_{\gamma 0}$, and amounts to 
$4 \%$ ($7 \%$) for
$E_\gamma > 10 \, MeV$ $(20 \, MeV)$.
\item The radiatively corrected decay width of the muon,
usually written as
\begin{equation}
\Gamma_\mu=\Gamma_0 \left[ 1+\frac{\alpha}{\pi}
(\frac{25}{8}-\frac{\pi^2}{2}) \right] \nonumber
\end{equation}
can be regarded as a sum of two mutually exclusive he\-li\-ci\-ty 
con\-tri\-bu\-tions~\cite{Fischer:2002hn}
\begin{eqnarray}
\Gamma_\mu=\Gamma_\mu(e_L^-)+\Gamma_\mu(e_R^-) \nonumber
\end{eqnarray}
where
\begin{eqnarray}
\Gamma_\mu(e_L^-) & = & \Gamma_0 \left[ 1+\frac{\alpha}{\pi}
(\frac{23}{8}-\frac{\pi^2}{2}) \right] \nonumber \\
\Gamma_\mu(e_R^-) & = & \Gamma_0 \, \frac{\alpha}{\pi} \, \frac{1}{4} \nonumber .
\end{eqnarray}
\item A full analysis of 
$\mu^- \to e^- \bar{\nu}_e \nu_\mu \gamma$,
aimed at finding regions of phase space with enhanced concentration
of right-handed electrons will be reported elsewhere.
\end{enumerate}

Acknowledgement: I wish to thank Volker Schulz for discussions, 
and for help in the preparation of this manuscript.

\end{document}